\documentclass[aps,epsfig,showpacs,showpacs,preprintnumbers,amsmath,amssymb,showpacs,10pt]{revtex4}
\usepackage[colorlinks=true, pdfstartview=FitV, linkcolor=red, citecolor=blue, urlcolor=blue]{hyperref}

\usepackage{amsmath,amsfonts,amssymb,graphics,graphicx,epsfig,color,times,bbm}
\usepackage{graphicx}
\usepackage{epstopdf}

\begin{document}

\title{\Large Multiple scattering and electron-uracil collisions \\
at low energies}

\author{S. Yalunin}
 \affiliation{Theoretical Physics Department, Kant State University
 (former Kaliningrad State University), \\
 A. Nevsky st. 14, Kaliningrad, Russia}
 \email{yalunin@bk.ru}

\author{S.B. Leble}
 \affiliation{
 Theoretical Physics and Mathematical Methods Department,
    Gda\'nsk University of Technology, ul, Narutowicza 11/12, Gdansk,  Poland}
 \email{leble@mifgate.pg.gda.pl}

\date{\today}

\begin{abstract}
Links between two well known methods: methods of zero-range and
non-overlapped (muffin-tin) potentials are discussed. Some difficulties of the
method of zero-range potentials and its possible elimination are
discussed. We argue that such advanced method of ZRP potential can be
applied to realistic electron-molecular processes. The method
reduces electron-molecule scattering to generalized eigenvalue
problem for hermitian matrices and admit fast numerical scheme. A
noteworthy feature of the method is direct possibility to calculate
the wave functions (partial waves). The theory is applied to
electron-uracil scattering. Partial phases and cross-sections at low
energies are evaluated and plotted.
    \center{Key words: electron-molecule
    scattering, uracil, multiple scattering,
    zero-range potential}
\end{abstract}

\maketitle

\section{Introduction}
\large
A growing interest to electron - big  molecules (particularly
- DNA and RNA bases) scattering  leads to a development of
one-center expansion of molecular potential methods
\cite{Giant2004,Tonzani}. However, such expansion seems to become
non-effective while molecular dimension grows: a molecular potential
may have many local minima which at the one-center expansion leads
to extremely non-smooth one-center radial matrix potential.  Such
difficulty does not appear within the method of non-overlapped
potentials (NP).

The notion of the non-overlapped potentials (named as the muffin-tin ones) appears in a context of
solid state theory \cite{John66} and, further, in scattering theory;
it exploits the idea of division of a molecule area into regions - atomic
spheres, which surround each atom, such that originally these spheres
touch but do not overlap. Inside each of the atomic spheres we
replace exact potential by a spherical averaged. Outside the
spheres, we can replace the potential by a constant potential
(ordinary taken for convenience to be zero). This concept is useful
in treatment of both polyatomic molecules and solids. For molecules,
this approximation gives a model for calculating electronic
energy levels \cite{Slat72} and scattering phases \cite{Zeische76}.
For crystals the application of the Bloch conditions instead of the
boundary condition of scattering theory leads directly to the
Korringa-Kohn-Rostoker formalism \cite{KKR}.

There is a special case of the non-overlapped potentials in which
radius of the atomic spheres is zero - zero-range potentials (ZRP).
The method of ZRP was proposed in 1936 by E. Fermi \cite{Fermi1936}
and has wide applications in photodetachment \cite{Arm1963},
inelastic scattering \cite{OU1983}, and other problem of the quantum
physics \cite{DO1975}. In such applications a potential is
represented by a boundary condition on a wave function and acts to
s-states. There are also generalizations to higher orbital angular
momentum states \cite{AR1977,ANDR}. To our knowledge none of these
generalizations have not applied to real molecular processes. It is
interesting to characterize some of the sources of the method
disability: (i) in realistic situation a choice of the potentials
parameters is very complicated (ii) eigenfunctions for ZRP are
singular and sometimes it is not clear how to avoid ambiguity in the
treatment of this singularity (iii) it is not clear how to improve
results of computations when necessary. In this paper we show that
ZRP method (without adjusted parameters), which include some
positive features of NP, can cope with realistic processes.

In the section II we show links between zero-range and
non-overlapped potentials and discuss how to avoid the problems. We
argue that the parameters in generalized boundary condition at an
atom can be obtained by solving one-dimension Schr\"{o}dinger
equation. If parameters is known, then partial phases and
cross-sections is readily obtained via matrix eigenvalue problem
\cite{ANDR}.

In the section III we consider an application - low energy
electron-uracil (U) scattering. We explain how to extract ZRP
parameters from the information provided by a quantum chemistry
package. We also summarize the computational steps involved in our
calculation and present results of our calculations: partial phases,
total and partial cross-sections.

We conclude with a short summary and perspectives.

\section{THEORY}
As it was mentioned in Introduction, we will consider a polyatomic
molecule as a system of spherically symmetric NPs, which we denote
as $V_i(|{\bf r}-{\bf a}_i|)$. It is convenient to write the
partial wave $\Psi({\bf r})$ as the linear combination
\begin{equation} \label{psi}
\Psi({\bf r}) = \sum_{i=1}^N\sum_{l=0}^{\infty}\sum_{m=-l}^{l}
x_{ilm} \Phi_{ilm}({\bf r}-{\bf a}_i)
\end{equation}
of the atomic waves $\Phi_{ilm}({\bf r})$
$$
\Phi_{ilm}({\bf r})=i^l \psi_{ilm}(r) Y_{lm}(\hat{\bf r}),
$$
where $i,l,m$ -- atom number and angular momentum quantum numbers
correspondingly; $\psi_{ilm}$ -- real functions of the radial
variable $r$; $Y_{lm}(\hat{\bf r})$ -- spherical harmonics
\cite{Varshal}; ${\bf a}_i$ -- the vector that marks the point of
expansion for each atom; $k$ -- radial wave number; $N$ -- number of
atoms in the molecule. The coefficient $x_{ilm}$ controls the
contribution of the wave $\Phi_{ilm}$ to the partial wave. We
suppose that outside the potential $V_i(r)$ the basis functions
$\{\Phi_{ilm}({\bf r})\}$ satisfy the Helmholtz equation
\cite{M1968}. Then outside the NP action the function $\psi_{ilm}$
must be the linear combination $c_2h_l^{(2)}(kr)-c_1h_l^{(1)}(kr)$
of the outgoing and ingoing spherical Bessel functions \cite{Abr}.
For a partial wave, the coefficients $c_{1,2}$ differ in the phase
factor $\exp(2i\delta)$, where $\delta$ is a phase shift. The
mentioned choice reflects the condition of equality of the fluxes of
the ingoing and outgoing waves at infinity. Without loss of
generality we may assume that the function $\psi_{ilm}$ is
normalized, so that at infinity
\begin{equation} \label{as}
\psi_{ilm}(r) = \cos\delta\, j_l(kr)-\sin\delta\, y_l(kr),
\end{equation}
where (we used property 10.1.1 of Ref. \cite{Abr}) $j_l(x)$ and
$y_l(x)$ are regular and singular spherical Bessel functions. To
derive differential equation for the functions $\psi_{ilm}(r)$ we
expand the partial wave at the vicinity of the point ${\bf a}_i$ in
the form
$$
\Psi({\bf r}_i)=\sum_{lm}x_{ilm} i^l
\left(\psi_{ilm}(r)+w_{ilm}j_l(kr)\right)Y_{lm}(\hat{\bf r}),
$$
where ${\mathbf r}_i={\mathbf r}+{\mathbf a}_i$, and substitute the
result to Schr\"{o}dinger equation. The result can be written as
inhomogeneous Schr\"{o}dinger equation for the function $\psi_{ilm}$
\begin{equation} \label{eq}
\left(H-\frac{_1}{^2}k^2\right)\psi_{ilm}( r)= -w_{ilm} V_i(r)
j_{l}(kr),
\end{equation}
where the coefficients $w_{ilm}$ describe the contribution from
waves that come from the other NPs, and hamiltonian $H$ is given by
the equation
$$
H=-\frac{1}{2}\left(\frac{d^2}{dr^2}+\frac{2}{r}\frac{d}{dr}-\frac{l(l+1)}{r^2}\right)+V_i(r).
$$

\begin{figure}[!h]
  \begin{center}
  \leavevmode
\resizebox{8 cm}{!}{\includegraphics{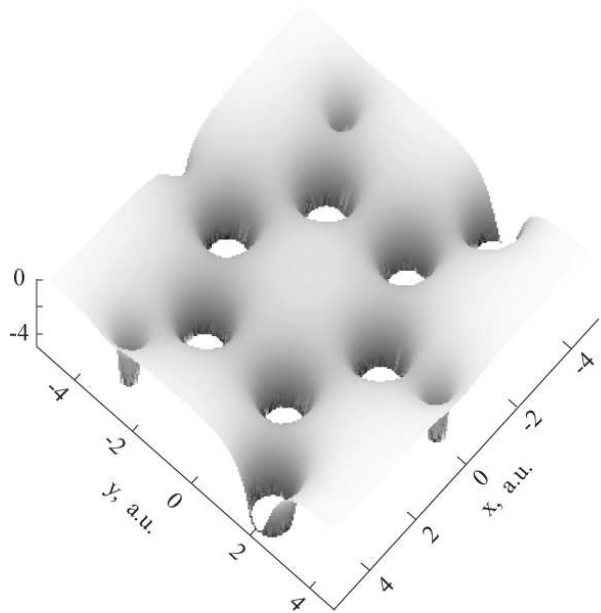}}
\end{center}
\caption{\label{f0} Electrostatic potential for ground state of
uracil as function of $x,y$. Potential was computed in ab initio
calculation and plotted in Hatree units.}
\end{figure}

Using the properties of the Helmholtz equation solutions
\cite{Varshal} one can show that the coefficients $w_{ilm}$ satisfy
the equation
\begin{eqnarray} \label{eqw}
w_{ilm}x_{ilm} = \sum_{j (i\neq j)} \sum_{l_1m_1} \sum_{l_2m_2}
x_{jl_1m_1} i^{l_2}Y_{l_2m_2}(\hat{\bf a}_{ij}) \nonumber
\\
\times Q^{l_1m_1}_{l_2m_2lm} \left(\cos\delta\, j_{l_2}(k|{\bf
a}_{ij}|)- \sin\delta\, y_{l_2}(k|{\bf a}_{ij}|)\right),
\end{eqnarray}
where we denote ${\bf a}_{ij}={\mathbf a}_i-{\mathbf a}_j$; symbol
$Q^{l_1m_1}_{l_2m_2lm}$ denotes the integral of three spherical
harmonics, multiplied by  $4\pi$. The upper indices mean that the
spherical function with indices $l_1,m_1$ should be complex
conjugate. This integral can be expressed in terms of Klebsh-Gordon
symbols, or, alternatively, via the $3jm$ Wigner symbols
\cite{Varshal}. In particular, by the Wigner-Eckart theorem the
integral is proportional to Clebsh-Gordan coefficients \cite{M1968}:
$$
Q^{l_1m_1}_{l_2m_2 lm} =\sqrt{4\pi \frac{(2l_2+1)(2l+1)}{(2l_1+1)}}
C^{l_10}_{l0l_20} C^{l_1m_1}_{lml_2m_2}.
$$
We can see that these symbols are real. From the coefficients
properties \cite{Varshal} it follows, that the nonzero symbols
correspond to the selection rule $m_1=m+m_2$, therefore the sum over
$m_2$ may be performed explicitly. While account the big $l$ values
one can use the selection rule: $l+l_1+l_2$ - even.

\begin{figure}[!h]
  \begin{center}
  \leavevmode
\resizebox{8 cm}{!}{\includegraphics{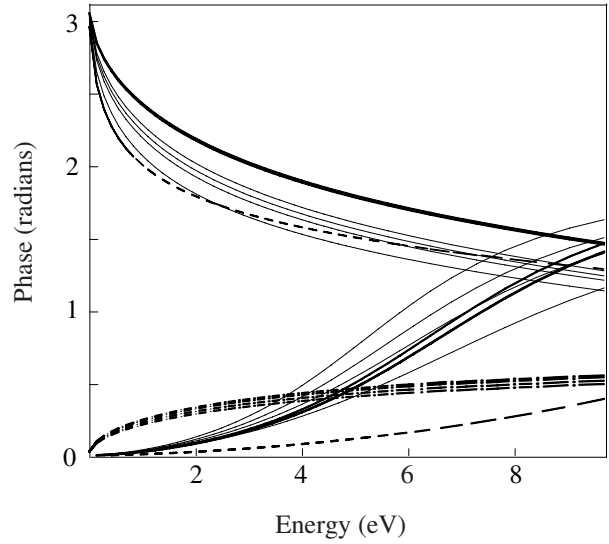}}
\end{center}
\caption{\label{f1} The scattering phases $\delta_{il}(E)$ for
different atoms and angular momenta $l=0,1$ as functions of incident
electron energy. Solid lines: carbon; dashed lines: oxygen; bold
lines: nitrogen; dash-dotted lines: hydrogen.}
\end{figure}

Note that, the function $\psi_{ilm}(r)$ does not coincide with the
radial component of the partial wave for $V_i(r)$, because of the
nonzero right-hand side of the equation $(\ref{eq})$. This means
that, in general, molecular partial shifts $\delta$ does not
coincide with partial shifts $\delta_{il}$ for a NP. However,
partial phases $\delta_{il}$, coefficients $w_{ilm}$, and molecular
partial shifts $\delta$ are linked by the equality
$$
w_{ilm}= \sin \delta (\cot\delta_{il}-\cot\delta).
$$

Combining this expression with $(\ref{eqw})$ we see, that the
unknown $x_{ilm}$ satisfy the system of linear algebraic equations,
which coincide with result of work \cite{ANDR}. In actual
calculations values of $l$ are always restricted. If the maximum
quantum number for all atoms are the same and equal to $L$, we have
the system of $N(L+1)^2$ equations, which gives the same number of
partial waves.

In practice it is convenient to work with matrices and column
vectors. Denote the column vector composed of the coefficients
$x_{ilm}$ as $|x\rangle$. Next define the matrix ${\sf S}$, so that
\begin{equation}\label{matrS}
{\sf S}_{ilm,jl_1m_1}= \sum_{l_2=|l_1-l|}^{l_1+l} i^{l_2}
Q^{l_1m_1}_{l_2m_2lm} j_{l_2}(k|{\bf a}_{ij}|) Y_{l_2m_2}(\hat{\bf
a}_{ij}),
\end{equation}
where $m_2=m_1-m$, and matrix ${\sf N}$
\begin{equation}\label{matrN}
{\sf N}_{ilm,jl_1m_1}=\sum_{l_2=|l_1-l|}^{l_1+l} i^{l_2}
Q^{l_1m_1}_{l_2m_2lm} y_{l_2}(k|{\bf a}_{ij}|) Y_{l_2m_2}(\hat{\bf
a}_{ij})
\end{equation}
for the case when $i$ and $j$ are different, and in the case $i=j$
nonzero elements are given by ${\sf N}_{ilm,ilm}=\cot\delta_{il}$.
The numbers $ilm$ and $jl_1m_1$ play the role of the multi-indices.
It is seen from the last expressions that, the matrix ${\sf S}$ is
positive definite and symmetric with respect to transposition of $j
l_1m_1$ and $ilm$, made simultaneously with the operation of complex
conjugation (hermicity). The matrix ${\sf N}$ is also hermitian. The
system for the unknown $x_{ilm}$ in the matrix form looks as the
eigenvalue problem
\begin{equation}\label{ep}
{\sf N}|x\rangle=\lambda {\sf S}|x\rangle
\end{equation}
for the eigenvalues $\lambda=\cot\delta$, which are real.

A result of the eigenvalue problem $(\ref{ep})$ solution is a set of
vectors  $|x\rangle$ and the partial phases $\delta$. If the
molecule has a symmetry, it is possible to classify the phases by
irreducible representations of the symmetry group. The multiplicity
of degeneration coincide with the dimension of the irreducible
subspace. Note, that partial phases, assumed as functions of energy,
only of different symmetry can intersect, while the intersection of
the partial phases of like symmetry is impossible.

Using known vectors $|x\rangle$ we can calculate the partial
harmonics
$$
A({\bf n})=\sum_{ilm}x_{ilm}\exp(-ik{\bf n}{\bf a}_i)Y_{lm}({\bf
n}).
$$
Partial harmonics of the molecular partial waves play the role of
spherical harmonics for the spherically symmetric potentials. In
analogy with the spherical harmonics, partial harmonics are
normalized with respect to integration over angles to the unit. This
means that vector $|x\rangle$ must be normalized, so that $\langle
x|{\sf S}|x\rangle=1$. The partial harmonics also may be applied to
calculation of the expectation value of angular momentum operator or
angular distribution of scattering for any partial wave. For
example, partial differential cross section has the form
\cite{DR1970}
$$
\frac{d\sigma}{d\Omega}=\frac{(4\pi)^2}{k^2} |A({\mathbf
n})|^2\sin^2\delta.
$$

The second step of the method is, if necessary, an evaluation of the
partial wave for the system of NPs. For this purpose the equation
$(\ref{eq})$ is integrated numerically with account the behavior at
infinity $(\ref{as})$, which is already known. After that the
partial wave is written as the linear combination $(\ref{psi})$ of
atomic waves $\Phi_{ilm}({\bf r})$. Further the perturbation theory
can be applied. Methods of perturbation theory (for example MP2
\cite{Rossi}) were found good for molecular states calculations, it
allows to expect to achieve good results by rather simple scheme.

\section{EXAMPLE of uracil}

As an illustration, we consider low energy electron-molecule
scattering for uracil molecules. The computational steps involved in
our calculation can be summarized as follows.

\begin{table}
\caption{\label{tab1} In the second column corresponding atomic
radii ($d_i$) are presented. In the third and fourth columns
$x_i,y_i$ are coordinates of the position vectors ${\bf a}_i$. We
also assume $z_i=0$.}
\begin{ruledtabular}
\begin{tabular}{lrrr}
Atom & $d_i$ & $x_i$ & $y_i$ \\ \hline
C1   & 1.37 & -2.29558  &  0.66988  \\
N2   & 1.23 & -2.14780  & -1.91792  \\
C3   & 1.37 &  0.11224  & -3.21127  \\
C4   & 1.37 &  2.32379  & -2.01273  \\
C5   & 1.37 &  2.37055  &  0.73276  \\
N6   & 1.23 &  0.00000  &  1.86414  \\
H7   & 0.64 & -0.03767  & -5.22343  \\
H8   & 0.64 &  4.08399  & -2.98680  \\
H9   & 0.64 & -3.78218  & -2.82664  \\
H10  & 0.64 & -0.04204  &  3.74037  \\
O11  & 0.93 & -4.30756  &  1.79697  \\
O12  & 0.93 &  4.27587  &  2.02817  \\
\end{tabular}
\end{ruledtabular}
\end{table}

(i) \emph{Choice of target basis and eigenstates.} We restrict our
consideration to the fixed-nuclear approximation and assume that
ground molecular state remains unperturbed during the scattering.
The single-determinant molecular wave function (in basis set
6-31G(d)) and static potential were obtained in ab initio
calculation with quantum chemistry package.

(ii) \emph{Choice of effective radii.} To compute NPs we divide all
space to non-overlapping areas $|{\bf r} -{\bf a}_i|\leqslant d_i$
corresponding to separate atoms with effective radii $d_i$.  In
order to evaluate the atomic radii $d_i$ we used the bond lengths,
which were obtained using a geometry optimization procedure, and the
following equations
$$
d_{i}+d_{j}=|{\bf a}_i-{\bf a}_j|.
$$
The effective radii are listed in Table \ref{tab1}.

(iii)\emph{ Choice of NPs.} To avoid the computational complexity
related with coupling terms, we average static interaction over
angles and compute the radial NPs which were expressed in the
following form
$$
V_{\textrm{st}}(r)= -\frac{q(r)}{r},\hspace{8mm}r \leqslant d
$$
where $q(r)$ denote effective charges. Our calculations show that
$q(r)$ is smooth function. Calculated static potential are plotted
in Figure $\ref{f0}$. To include the nonlocal exchange interaction
we used local density approximation, originally proposed by Slater
\cite{Slater1951,Slater1960} for atomic problems. Further the
approximation of the local potential was applied to scattering
problems \cite{Hammer1957,Hara1967,Tonz2004} and calculating
molecular quantum defects for small closed-shell target molecules
\cite{Tashiro}. In the paper of Hara \cite{Hara1967} the following
effective potential, depending on energy of impact electron is
introduced
\begin{equation} \label{pHara}
V_{\textrm{ex}}(r)=-\frac{2}{\pi}k_{\textrm{F}}(r)\left(
\frac{1}{2}+\frac{1-\eta^2}{4\eta}\ln\left|
\frac{1+\eta}{1-\eta}\right| \right),
\end{equation}
where $k_{\textrm{F}}(r)=[3\pi^2\rho(r)]^{1/3}$ - Fermi momentum,
and $\rho(r)$ denotes averaged (over angles around the atom)
electron density. These densities were obtained in ab initio
calculation via density matrix. The parameter $\eta$ is given by the
equality
\begin{equation} \label{eta}
\eta =[k^2+2I+k_{\textrm{F}}^2(r)]^{1/2}/k_{\textrm{F}}(r),
\end{equation}
where $I$ - molecule ionization energy (in a.u.). The equation
$(\ref{eta})$ emerges from the assumption that the scattered
electron and the electron in the highest energy bound state (i.e.
Fermi electron, which has momentum $k_{\textrm{F}}(r)$) move in the
same potential field. For the uracil molecule the experimental value
of the ionization energy is estimated as $I=8.35$ eV. The exchange
Hara's potential $(\ref{pHara})$ is attractive  (at arbitrary $k$),
because the expression in brackets is always positive. When the
scattering electron energy ($k^2/2$) grows, the expression at the
bracket uniformly decrease, that qualitatively in accordance with
the these about decreasing with energy contribution of exchange
integral.

\begin{figure}
  \begin{center}
\resizebox{8 cm}{!}{\includegraphics{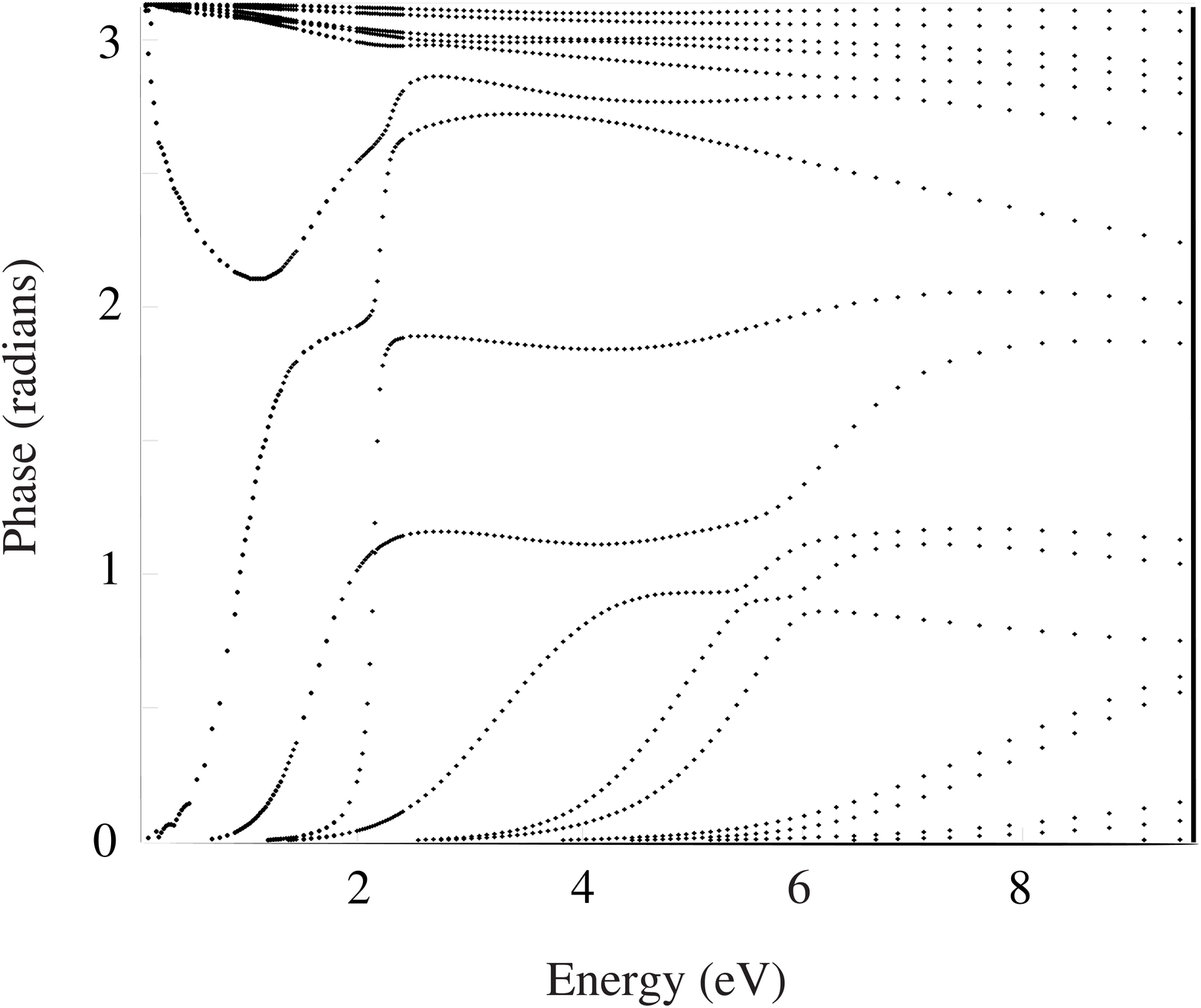}}
\end{center}
\caption{\label{f2a} Computed partial phases $\delta(E)$ for $A'$
symmetry as functions of incident electron energy.}
\end{figure}

\begin{figure}
  \begin{center}
\resizebox{8 cm}{!}{\includegraphics{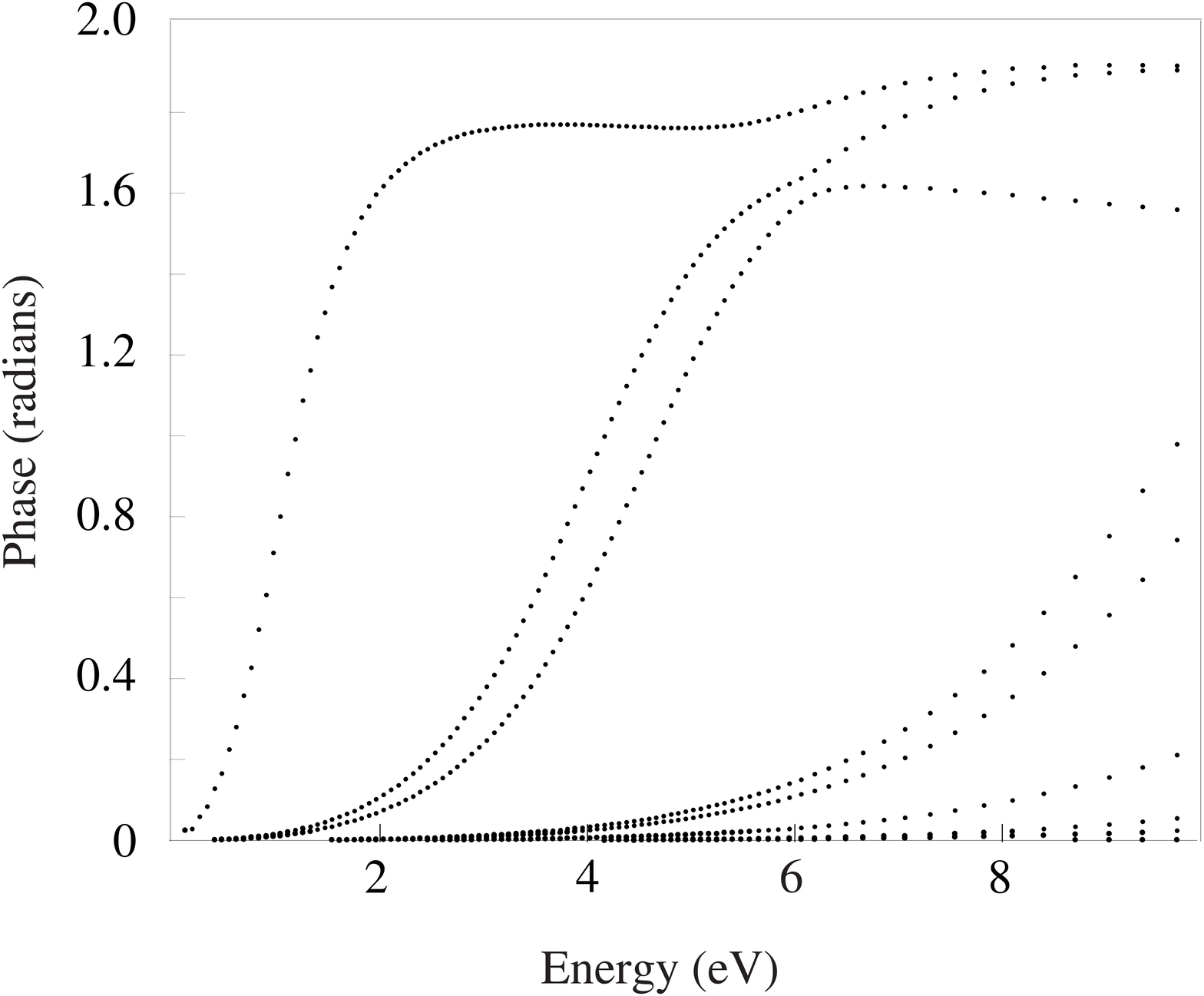}}
\end{center}
\caption{\label{f2b} Computed partial phases $\delta(E)$ for $A''$
symmetry as functions of incident electron energy.}
\end{figure}

(iv) \emph{Computation of partial shifts.} To obtain the partial
shifts $\delta_{il}$, we solved numerically the radial
Schr\"{o}dinger equation with the potentials, which were obtained
from step (iii). The partial shifts were calculated for angular
momenta $l=0,1$ and are ploted in Figure $\ref{f1}$.

(v) \emph{Computation of the matrices ${\sf S}$ and ${\sf N}$.
}These matrixes (for every irreducible representations) were
constructed by the formulas $(\ref{matrS})$ and $(\ref{matrN})$.

\begin{figure}
  \begin{center}
\resizebox{8 cm}{!}{\includegraphics{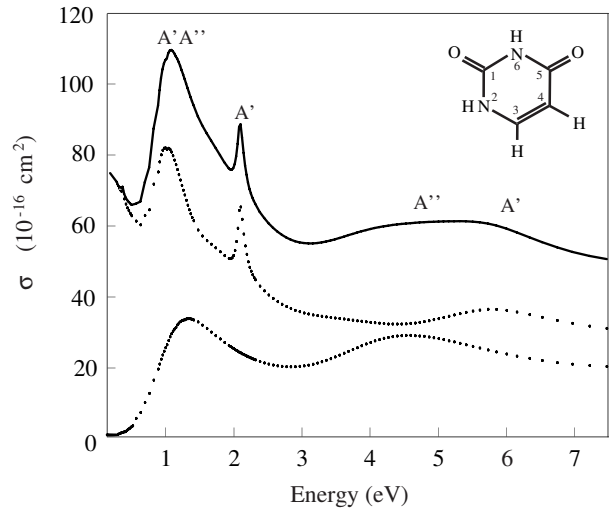}}
\end{center}
\caption{\label{f3} Cross sections as functions of incident electron
energy. Solid line: our integral cross sections; upper points:
partial cross sections with the symmetry $\text{A}'$; lower points:
partial cross sections with symmetry $\text{A}''$. }
\end{figure}

(vi)\emph{ Computation of $\delta$ and $\sigma$.} Using the matrix
elements assembled from step (v), the molecular partial shifts were
calculated from eigenvalue problem $(\ref{as})$ using Cholesky
decomposition. The results of the computations at different energies
are summarized in Figures $\ref{f2a}$ and $\ref{f2b}$. The integral
cross section, averaged over molecular orientations, were found by
formula $\sigma=(4\pi/k^2)\sum \sin^2\delta$. The integral cross
sections are plotted in Figure $\ref{f3}$.

(vii) \emph{Computation of resonances.} The resonance positions and
widths were found via the equations
$$
\delta(E)=\pi/2, \hspace{6mm} \Gamma = 2 \delta'(E)^{-1}
$$
and are listed in Table \ref{tab2}.

\begin{table}
\caption{\label{tab2} Resonance parameters.}
\begin{ruledtabular}
\begin{tabular}{lcc}
Symmetry & $E_{\text{res}}$     & $\Gamma$ \\
      & (eV)      & (eV)    \\    \hline
$A'$ & 1.2  & 1.1  \\
     & 2.2  & 0.3  \\
     & 6.7  & 1.5  \\
$A''$& 1.9  & 4.4  \\
     & 5.6  & --   \\
     & 6.1  & --   \\
\end{tabular}
\end{ruledtabular}
\end{table}

The Figure $\ref{f3}$ shows that resonance positions (2.16, 5.16,
7.8 eV) found in $R$-matrix calculations \cite{Tonzani} agree with
our calculations. The sharp peak at 2.2 eV in Figure $\ref{f3}$ may
also arise from long-lived anion state, whose lifetime is
sufficiently long to allow nuclear motion and, as consequence,
possible fragmentation. Indeed, recent experiments
\cite{Hanel,Scheer} showed effective destruction of uracil through
dissociative electron attachment (DEA). Also a large peak around 1.0
eV was observed. In Ref. \cite{Scheer} authors explain that the
sharp peak, in their DEA measurements, can be associated with
vibrational Feshbach resonance, that is, exited vibrational levels
of the dipole bound anion states of these compounds. In should be
noted, that we exclude (for simplicity) dipole interaction. Thus,
our sharp peak indicates another possible explanation, which,
however, does not exclude the arguments of Ref. \cite{Scheer}.

\section{CONCLUSION}

We have discussed the "multiple-scattering" method emerged from the
zero-range and muffin-tin theories. The method reduces
electron-molecule scattering to generalized eigenvalue problem
$(\ref{ep})$ for hermitian matrices. A noteworthy feature of the
method is direct possibility to calculate the wave functions
(partial waves). This way is based on one-dimensional inhomogeneous
Schr\"{o}dinger equation $(\ref{eq})$ for radial component
$\psi_{ilm}(r)$ and only requires the atomic potential and partial
phase of given partial wave. Thus, this approach may be especially
important in relation to Bardsley-Fano theory \cite{Bard68,Fano61}
and dissociative attachment at low energies.

The preliminary results on uracil presented here are quite
promising. Particularly, our positions of the shape resonances
demonstrate good agreement with the $R$-matrix calculations
\cite{Tonzani}. Thus, this method can be applied to another DNA
bases, such as adenine, cytosine, thymine etc. In this connection,
it is interesting and important to extend the method to the case of
long range interactions, such as dipole interactions. Note, this
method gives fast numerical scheme and, therefore, we have good base
for future approximations and modeling of electron-DNA scattering in
sense of the works \cite{Car03}.

\end{document}